\documentclass[aps,prl,twocolumn]{revtex4}
\usepackage{graphicx}
\usepackage{dcolumn}
\usepackage{amsmath}
\usepackage{array}
\usepackage{epsfig}
\usepackage{amssymb}
\begin{document}

\title{Impact of the Spin Density Wave Order on the Superconducting Gap of Ba(Fe$_{1-x}$Co$_x$)$_2$As$_2$}
\author{L. Chauvi\`ere}
\author{Y. Gallais}
\author{M. Cazayous}
\author{M.A. M\'easson}
\author{A. Sacuto}
\affiliation{Laboratoire Mat\'eriaux et Ph\'enom\`enes Quantiques, UMR 7162 CNRS, Universit\'e Paris Diderot Paris 7, B$\hat{a}$t. Condorcet 75205 Paris Cedex 13, France}
\author{D. Colson}
\author{A. Forget}
\affiliation{CEA Saclay, IRAMIS, Service de Physique de l'Etat Condens\'e (SPEC CNRS URA 2464), F-91191 Gif-sur-Yvette, France}

\begin{abstract}
We report a doping dependent electronic Raman scattering measurements on iron-pnictide superconductor Ba(Fe$_{1-x}$Co$_x$)$_2$As$_2$ single crystals. The B$_{2g}$ Raman spectrum at optimal doping is consistent with a strongly anisotropic gap on the electron pocket. Upon entering the coexistence region between superconducting (SC) and spin-density-wave (SDW) orders, the effective pairing energy scale is strongly reduced. Our results are interpreted in terms of a competition between SC and SDW orders for electronic states at the Fermi level. Our findings advocate for a strong connection between the SC and SDW gaps anisotropies which are both linked to interband interactions.
\end{abstract}

\maketitle

The competition and/or coexistence between different electronic orders is a central issue in the physics of strongly correlated systems and in particular in the physics of unconventionnal or electronic driven superconductivity. 
%This was illustrated early on in dichalcogenide superconductors displaying both charge density wave (CDW)  and superconducting (SC) orders like in NbSe$_2$ \cite{So}, or in heavy fermions superconductors where magnetic and SC orders are intimately related \cite{Mathur}.
%In the latter, the magnetic fluctuations are believed to serve as a glue for superconductivity. 
%Very often however it is not immediatly clear whether or not the competing order is favouring or not superconductivity. 
The recently discovered iron-pnictides high temperature superconductors provide an interesting case to study the coexistence and/or competition between spin-density-wave (SDW) and SC orders. The delicate balance between both orders in the pnictides is exemplified by two facts: superconductivity in iron-pnictides only arises when the SDW order is significantly weakened, and yet there is a growing theoretical consensus that both orders are driven by interband interactions \cite{Mazin,Kuroki1,Wang}. The predominance of interband interactions naturally leads to a superconducting order parameter that switches sign between different electronic bands \cite{Mazin,Kuroki1}, which contrary to a regular s-wave gap, favors the coexistence between SDW and SC orders \cite{Vorontsov,Fernandes-long}. 
%The interplay between magnetism and superconductivity is particularly delicate in pnicitides because of the itinerant nature of magnetism, implying that SC and SDW orders essentially share the same conducting electrons \cite{Vorontsov,Fernandes-long}. 

\par
Electron doped Ba(Fe$_{1-x}$Co$_x$)$_2$As$_2$ (Co-Ba122) is a particularly suitable system to study the interplay between both orders because Co doping can be used to tune SDW and SC orders presumably by changing the respective size of the Fermi surface sheets and the corresponding nesting properties. In addition, there are strong evidences that both SDW and SC orders coexist spatially over a finite range of doping \cite{Laplace}. Recent neutron scattering measurements have shown a sizable reduction of the Fe magnetic moment upon entering the superconducting state that most likely results from a competition between the two orders for low energy electronic states \cite{Neutrons,Fernandes}. Up to now however, the impact of the SDW order on the superconducting properties themselves such as the gap amplitude and anisotropy has not been experimentally addressed.

\par
Here we report doping dependent electronic Raman scattering measurements on Ba(Fe$_{1-x}$Co$_x$)$_2$As$_2$ (Co-Ba122) single crystals in the superconducting and normal states. In optimally doped and underdoped crystals, the superconducting spectra show a clear two-component response, $2\Delta_{max}$ and $2\Delta_{min}$, that is interpreted in terms of an anisotropic s-wave gap. The superconducting gap is strongly renormalized upon entering SDW-SC region:  the $2\Delta_{max}$ component intensity is strongly suppressed, leaving a weaker $2\Delta_{min}$ pair-breaking peak at much lower energy. We propose a picture of the SC-SDW coexistence where the SDW orders effectively gaps out part of the Fermi surface preventing SC order but leaves other parts essentially unaffected where SC can emerge.

\par
Single crystals of Co-Ba122 were grown from a self-flux method as described elsewhere \cite{Colson}. 
%Crystals from the same batches were also studied by transport  \cite{Colson}, NMR \cite{Laplace}, M$\ddot{o}$ssbauer spectroscopy \cite{Bonville} and ARPES \cite{Brouet} measurements.
 Samples were freshly cleaved and cooled in a $^4$He cryostat with a base temperature of 2.7~K. The spectra reported here were performed using the 514.5$~nm$  and 647$~nm$ lines of an Ar-Kr laser. All the temperatures reported here take into account the estimated laser heating \cite{Chauviere}. The scattered light was analyzed by a triple grating spectrometer equipped with a liquid nitrogen cooled CCD detector. Samples orientation was adjusted so that the crossed polarizations are along the diagonals of the Fe-square planes. With this configuration, we probe the B$_{2g}$ symmetry in all Raman spectra shown here.

\begin{figure}
	\centering
	\epsfig{figure=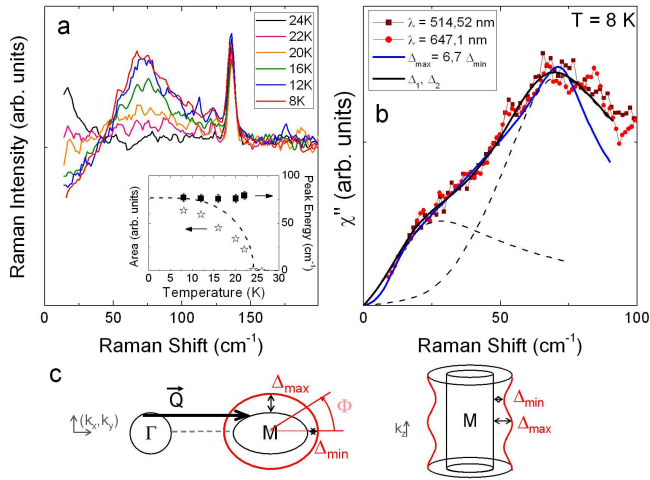, width=1\linewidth, clip=}
	\caption{Color online. a) Evolution of Raman intensity with temperature in the $B_{2g}$ symmetry for optimal doping $x = 0.065$ - Inset : Integrated area and Energy of the pair-breaking peak as a function of temperature. The dashed line is the BCS temperature dependence of 2$\Delta$. b) Raman response $\chi''\sim(1+n(\omega,T))^{-1} \times I$, where $n(\omega,T)$ is the Bose factor and $I$ is the measured Raman intensity,  for different laser wavelengths $\lambda = 514.52$ and $647.1$~$nm$ at $T = 8 K$. The theoretical fit in blue was obtained using an anisotropic gap (see text), a broadening $\gamma=0.1 \Delta_{max}$ and a constant B$_{2g}$ Raman vertex around the electron pocket \cite{Boyd}. The theoretical fit in black was obtained using two isotropic gaps $2\Delta_1, 2\Delta_2$(dashed lines), a broadening $\gamma=0.3 \Delta_2$. c) Sketches of an anisotropic s-wave gap around the M electron pocket in the (k$_x$,k$_y$) plane (left) or along k$_z$ (right).}
	\label{fig1}
\end{figure}

\par
Figure \ref{fig1}a displays the evolution of the Raman intensity in the $B_{2g}$ symmetry through the superconducting transition temperature for x=0.065 (T$_c$=24.7~K). Below T$_c$, a SC pair-breaking peak emerges around 75~cm$^{-1}$ yielding  $\frac{2\Delta_{max}}{k_bT_c} \approx 4.4$. The overall shape of the spectrum and the pair-breaking energy are in agreement with an earlier Raman report on a similarly doped crystal \cite{Hackl}. The extracted pair-breaking energy is also in agreement with the magnitude of the superconducting gap found for the electron pocket by ARPES \cite{Terashima}, indicating that the B$_{2g}$ Raman symmetry predominantly probes the M electron pockets \cite{Hackl}, and in agreement with band structure calculations of the Raman vertex \cite{Mazin-Raman}. The integrated area of SC peak  decreases continuously and disappears at $T_c$ while the peak energy itself does not show any sizable softening up to $T_c$ within our experimental accuracy (inset of Fig.\ref{fig1}a).

\par
In Fig.\ref{fig1}b is shown a zoom of the SC Raman response $\chi"$ below 100~cm$^{-1}$ and down to 8~cm$^{-1}$, at two different excitation wavelengths, $\lambda = 514.52~nm$ and $647.1~nm$. Both responses are essentially identical and display considerable spectral weight below the main pair-breaking peak with a weak downward bend below 20~cm$^{-1}$. The low energy part of the spectrum can be linked to an anisotropic s-wave gap around the M electron pocket, as drawn schematically in Fig.\ref{fig1}c. This picture is in agreement with thermal conductivity and penetration depth measurements where significant gap anisotropy was reported in Co-Ba122 \cite{Tanatar,Gordon} and is consistent with calculations based on spin-fluctuations mediated superconductivity \cite{Maier,Chubukov-Gap}. Reasonably good fits of the low energy spectrum are obtained using the standard BCS Raman response \cite{Klein} with a phenomenological broadening $\gamma$ and an in-plane anisotropy of the gap, $\Delta(\phi) = \Delta_{max} \frac{a + cos(2\Phi)}{a+1} $ with a = 1.35~($\pm$0.1) (see Fig.\ref{fig1}b) yielding deep minima in the gap function $\Delta_{max} = 7 ~(\pm 2)~ \Delta_{min}$. 

\par
An equally good fit was obtained by invoking a strong k$_z$ dependence of the gap (see Fig.\ref{fig1}c) instead of the in-plane anisotropy used above. Recent c-axis thermal conductivity data on Co-Ba122 suggest a strong k$_z$ dependence of the superconducting gap. However, this dependence most likely arises from the 3D hole Fermi surface centered around $\Gamma$ \cite{Reid} and not the 2D electron pocket which is probed in B$_{2g}$ symmetry. We note that because of the presence of two electron pockets in the reduced Brillouin zone, a two gaps scenario cannot be ruled out. Indeed, a satisfactory fit of the data could also be performed using two isotropic gaps $\Delta_1$, $\Delta_2$ with $\Delta_2 \sim 4 \Delta_1$ but with a rather large lifetime broadening, $\gamma=0.3\Delta_2$ (see Fig.\ref{fig1}b).

\begin{figure}
	\centering
	\epsfig{figure=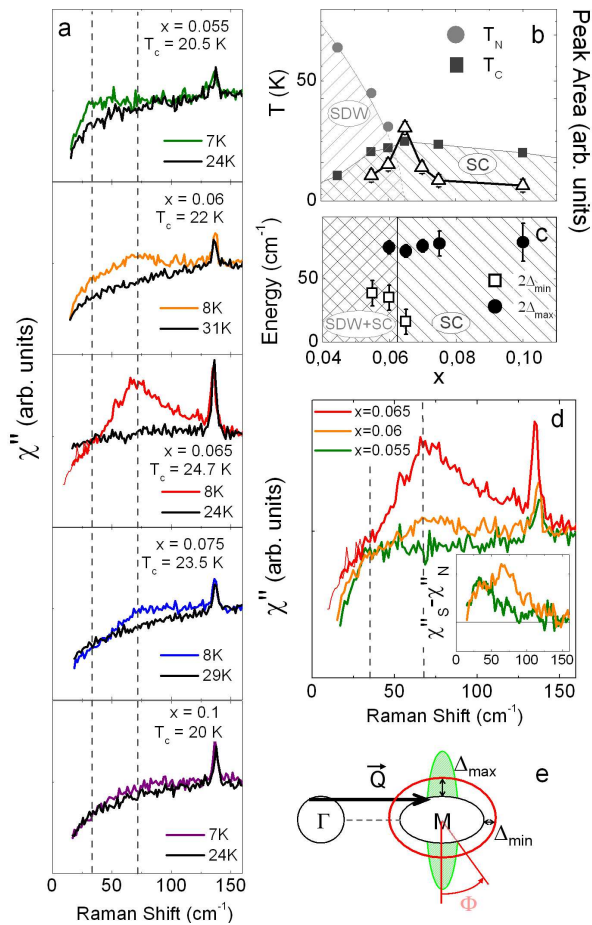, width=0.9\linewidth, clip=}
	\caption{Color online. a) Raman response $\chi''$ in the $B_{2g}$ symmetry above and below the SC transition for $x=0.055$, $x=0.06$, $x=0.065$, $x=0.075$ and $x=0.1$. b) Phase diagram of Co-Ba122 showing the superconducting peak area, defined as $\int_{15~cm^{-1}}^{125~cm{-1}}\chi"(\omega)d\omega$ (empty triangles) as a function of doping. The overall intensity of the spectra were normalized with the response above 200~cm$^{-1}$ which was found to be weakly doping dependent between x=0.055 and x=0.10. c) Doping evolution of the two component, $2\Delta_{max}$ and $2\Delta_{min}$ in an anisotropic gap picture. $2\Delta_{min}$ values are not reported for x=0.075 and x=0.10 because of the absence of a clear low energy component in the spectra. d) Raman response at low temperature for indicated dopings - Inset : substraction of $\chi''$ at $T = 8~K$ in the SC state and at $T = 30~K$ in the normal state for $x = 0.055$ and $x = 0.06$. e) Sketch of the competition between SDW and SC gaps on the $M$ electron pocket.}
	\label{fig2}
\end{figure}

\par
We now turn to the evolution of the superconducting response with varying doping and in particular across the SDW transition. While no magnetic transition is observed above T$_c$ for x=0.065 (T$_c$=24,7~K), 0.075 (T$_c$=23.5~K) and 0.10 (T$_c$=20~K), the x=0.06 (T$_c$=22~K) and x=0.055 (T$_c$=20.5~K) crystals display magnetic SDW transition (T$_N$) at 41~K and 31~K respectively (see the phase diagram in Fig.\ref{fig2}b) \cite{Laplace,Colson}. The B$_{2g}$ Raman responses, well below and slightly above T$_c$, are displayed in Fig.\ref{fig2}a as a function of doping. No pair-breaking peak was observed in B$_{1g}$ symmetry for all dopings ; weak and broad pair-breaking peaks were observed between 100 and 180~cm$^{-1}$ in the A$_{1g}$ channel \cite{Hackl}.

\par
All B$_{2g}$ responses display a distinctive pair-breaking peak in the superconducting state but its overall intensity varies significantly with doping. The weaker superconducting responses for x=0.075 and x=0.10 do not allow a reliable extraction of the gap anisotropy as performed in the x=0.065 case. However, the sizable intensities well below 2$\Delta_{max}$ are consistent with a significant gap anisotropy, possibly including nodes (see later for a discussion of the x=0.06 and x=0.055 cases). The doping dependence of the pair-breaking peak integrated intensity, which is proportionnal to the Cooper pair density in the BCS framework \cite{Blanc}, is sharply peaked at x=0.065 (see Fig.\ref{fig2}b). The suppression of the Cooper pair density away from optimal doping is in agreement with specific heat measurements which report a similar behavior for the specific heat jump across T$_c$ \cite{Budko}. This effect cannot be simply linked with disorder as one would naively expect because the residual resistivity extracted from transport measurements does not vary significantly over the corresponding doping range \cite{Colson}. In the same manner, it cannot be simply linked to the T$_c$ itself since the strong changes in Cooper pair density reported here correspond to relatively modest changes in T$_c$ (see for example x=0.065, T$_c$=24.7~K and x=0.075, T$_c$=23.5~K). Rather it is highly suggestive of a strong link between the Cooper pair density and the proximity of the critical doping where the SDW phase disappears. 

\par
The way the superconducting response is suppressed differs drastically between underdoped crystals, where SDW and SC coexist, and overdoped crystals with only the SC phase. While the SC response remains dominated by the main pair-breaking peak centered around 75~cm$^{-1}$ in the overdoped regime (x=0.075 and x=0.10), in the underdoped regime, x=0.06 and x=0.055, the overall spectral weight of the SC response becomes gradually localized at lower energies upon entering the SDW-SC phase (see Fig.\ref{fig2}a). This evolution does not result from a gradual decrease of the gap energy. Rather, the 2$\Delta_{max}$ pair breaking peak intensity is gradually suppressed when going from x=0.065 to x=0.055, leaving a weak but distinct shoulder at much lower energies ($\sim$ 35~cm$^{-1}$) that ultimately dominates the superconducting response for x=0.055 (see Fig.\ref{fig2}c and d). This is highlighted in the inset of Fig.\ref{fig2}d which shows the substraction of the Raman response between the SC state and just above $T_c$. The $x = 0.055$ response only contains a single component at 35~cm$^{-1}$ whereas it still shows two contributions, at 35~cm$^{-1}$ and 75~cm$^{-1}$, for $x = 0.06$ (see Fig.\ref{fig2}c). We stress that the drastic reduction of the apparent pairing energy scale cannot be ascribed to the lower T$_c$ since both x=0.055 and x=0.10 have similar T$_c$ and yet very different pair-breaking peak energies. Rather it is a direct consequence of the competition between SDW and SC phases. 
 
\par
In a simple view of the SDW formation, the nesting $\vec{Q}=(\pi, \pi)$ between the circular hole-pockets in the center of the Brillouin zone $\Gamma$ and the ellipsoïdal electron pockets in the corner of the Brillouin zone $M$ is imperfect \cite{Brouet,Liu}. The gap associated with the SDW formation is thus expected to open only on the nested parts of $M$ electron pockets (see Fig.\ref{fig2}e in green) preventing or strongly altering the formation of the superconducting phase in these regions of the Fermi surface \cite{Kato,Vorontsov,Fernandes}. Recent ARPES experiments do indeed show an anisotropic suppression of the ARPES intensity around the M electron pocket upon entering the SDW phase \cite{Liu}. Within this simple k-space SDW-SC coexistence picture, and assuming an anisotropic superconducting gap around the M electron pocket, our data imply that the SDW gap opens precisely where the SC gap is maximum, strongly suppressing the 2$\Delta_{max}$ peak, while leaving essentially unaffected the FS regions where the gap amplitude is smaller (i.e. close to 2$\Delta_{min}$, see Fig.\ref{fig2}e). A similar picture holds in a two gap model, where the SDW gap opens only on the most nested pocket, associated with the large SC gap $\Delta_2$, leaving unaffacted the other one, associated with the small SC gap, $\Delta_1$. In both cases, the strong connection between the SDW and SC gap amplitudes suggests that both phases are driven by interband Coulomb scattering \cite{Terashima,Wang}.

\begin{figure}
	\centering
	\epsfig{figure=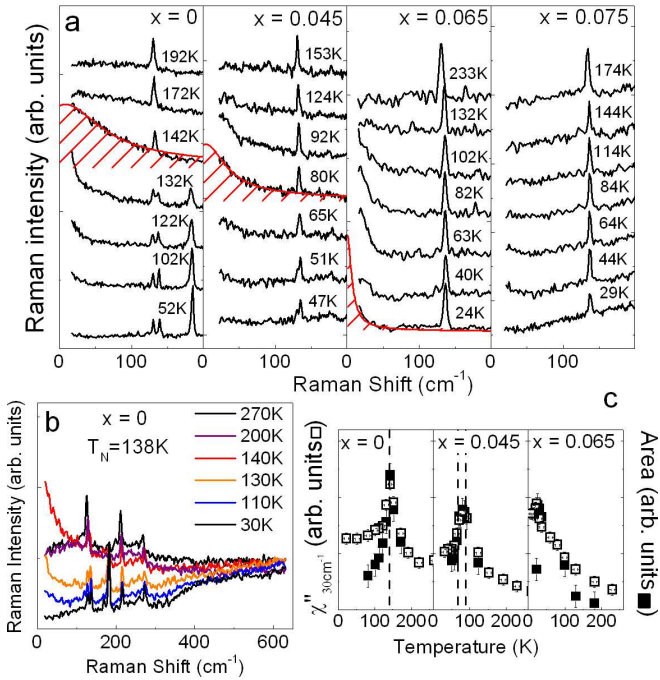, width=1\linewidth, clip=}
	\caption{Color online. a) Temperature dependence of the low energy Raman intensity in the $B_{2g}$ symmetry for $x=0$ (T$_N$=138~K), $x=0.045$ (T$_N$=64~K), $x=0.065$ and $x=0.075$. Several phonon anomalies are observed at the tetragonal to orthorhombic transition \cite{Chauviere}. b) Temperature dependence of the Raman continuum up to 600~cm$^{-1}$ for $x$=0. c) Area of the Lorentzian QEP (full squares, highlighted in red in panel a) and Raman response at 30~cm$^{-1}$ (empty squares) as a function of temperature. Structural and magnetic transitions are indicated in dashed lines.}
	\label{fig3}
\end{figure}

\par
The SDW transition has also a strong impact on the normal state Raman continuum. Figure \ref{fig3}a shows temperature dependent measurements in the normal state for different dopings $x=0$, $x=0.045$, $x=0.065$, $x=0.075$ in the $B_{2g}$ symmetry. The Raman continuum intensity displays a systematic increase at low energy around the magneto-structural transition temperature (shown in dashed lines in Fig.\ref{fig3}c) for the x=0 and x=0.045 crystals before disappearing at lower temperatures. It is also observed for the x=0.065 crystals where no magneto-structural transition is observed above T$_c$, but is essentially absent for x=0.075 and x=0.10 (not shown). A similar Quasi Elastic Peak (QEP) was observed in the Raman spectra of undoped Sr(FeAs)$_2$ \cite{Lemmens}. At higher energy and  for x=0, there is a strong suppression of the Raman intensity below 400~cm$^{-1}$ below T$_N$ which is linked to the opening of the SDW gap (see Fig.\ref{fig3}b) in agreement with optical conductivity meaurements \cite{Hu}. The suppression weakens at higher doping levels and becomes hardly detectable for $x$$\geqslant$0.045.

\par
The QEP was analyzed using a Lorenzian lineshape with I($\omega$)$\sim$(1+n($\omega$,T))$\frac{\omega\gamma}{\omega^2+\gamma^2}$ where $n(\omega,T)$ is the Bose factor and $\gamma$ the width at half-maximum. As shown in Fig.\ref{fig3}, the QEP intensity is maximum at the transition for x=0 and x=0.045 suggesting it originates from magnetic energy fluctuations as observed in magnetic insulators close to T$_N$ \cite{Reiter}. In the case of iron-pnictides, the presence of fluctuating magnetic domains above the transition could also give rise to the observed QEP \cite{Mazin2}.

\par
In conclusion, we have reported a doping dependent study of the superconducting gap in Ba(Fe$_{1-x}$Co$_x$)$_2$As$_2$. The effective pairing scale is strongly reduced upon entering the SDW-SC coexistence phase. It is interpreted as a consequence of the competition between the SC and the SDW orders. Our study illustrates the delicate interplay between magnetism and superconductivity which seems to be a generic feature of iron-pnictide superconductors. 

\par
We thank F. Rullier-Albenque for providing us with transport data. We gratefully acknowledge fruitful discussions with J. Bobroff, V. Brouet, I. Mazin, F. Rullier-Albenque and Y. Sidis.

\end{document}